# Magnetoplasmon-surface phonon polaritons coupling effects in radiative heat transfer

MINGJIAN HE,[1,2] HONG QI,[1,2,*] YATAO REN,[1,2] YIJUN ZHAO,[1] MAURO ANTEZZA [3,4,5]

[1] *School of Energy Science and Engineering, Harbin Institute of Technology, Harbin 150001, P. R. China*
[2] *Key Laboratory of Aerospace Thermophysics, Ministry of Industry and Information Technology, Harbin 150001, P. R. China*
[3] *Laboratoire Charles Coulomb (L2C), UMR 5221 CNRS-Université de Montpellier, F-34095 Montpellier, France*
[4] *Institut Universitaire de France, 1 rue Descartes, F-75231 Paris, France*
[5] *E-mail: mauro.antezza@umontpellier.fr*

*Corresponding author: qihong@hit.edu.cn*

**In this letter, based on the quantum Hall regime of magneto-optical graphene, we have theoretically investigated the coupling of magnetoplasmon polaritons (MPP) to surface phonon polaritons (SPhPs) by investigating the radiative heat transfer between two graphene-coated SiO₂ slabs. By applying an external magnetic field, the separated branches of intraband and interband MPP can both couple with SPhPs to form tunable modes, which remould the energy transport of the system. The heat transfer mechanism is completely changed from enhancement to attenuation due to the strong coupling, and the thermal stealthy is realized for the graphene. The letter has great significance for the graphene-based magneto-optical devices.** © 2019 Optical Society of America

http://dx.doi.org/10.1364/OL.99.099999

Near-field radiative heat transfer (NFRHT) can be greatly enhanced at the nanoscale and much larger than the Planck's blackbody limit [1], especially when the surface phonon polaritons (SPhPs) [2-4] or surface plasmon polaritons (SPPs) [5-7] are excited. Graphene, as an intriguing two-dimensional material, has extensive advantages over other film materials, e.g., the tunable carrier concentration [8], strong interactions with light [9], and enhanced optoelectronics with plasmons [10]. Owning to excellent optoelectronic performance and strong SPPs, in the field of NFRHT, numerous unique phenomena are discovered between graphene sheets [11], gratings [5, 6, 12, 13], disks [14, 15] and multilayer systems [7, 16].

With the application of an external magnetic field, considerable cyclotronic energies are gained by the electrons in graphene via the Lorentz force, hybridization between cyclotron excitations and plasmons occurs, originating magnetoplasmon polaritons (MPP) [17]. The continuum Dirac energy spectrum converts into separated Landau levels (LLs) with the energies of $n$-th LL given by $E_n = \text{sign}(n)(\hbar v_F / l_B)\sqrt{2|n|}$, where $v_F \approx 10^6$ m/s is the Fermi velocity of the carriers in graphene, $l_B = \sqrt{\hbar/(eB)}$ is the magnetic length, and $B$ is the intensity of the magnetic field. Induced by the magneto-optical characteristics of graphene [18], some unusual phenomena occur such as quantum Faraday effects and Kerr effects [19, 20]. Very recently, in suspended magneto-optical graphene, Shubnikov–de Haas-like radiative heat flux is observed, caused by the energy transitions between various LL*s* [21, 22]. Then using twisted graphene gratings, the combined effect of grating and magnetic field is utilized to realize MPP manipulation of NFRHT based on the quantum Hall regime of graphene [12]. In the absence of magnetic fields, the graphene SPPs can interact with the SPhPs of polar materials to form strongly coupled modes to tune the NFRHT [5, 23-25]. Nevertheless, the existing researches focus only on the isolated effect of MPP on NFRHT, but the coupled effects of MPP and SPhPs are ignored, which have great significance in the laboratory physics when graphene is always attached to the substrate [26]. The interactions and coupling between them should have great influence on the NFRHT, for the reason that the hybridization between magnetoplasmon and the polar optical phonon modes can couple resonantly with electromagnetic fields.

In this letter, in the presence of magnetic fields, the coupled effects of MPP and SPhPs on the NFRHT are investigated, which provides practical significance for fundamental research and new applications in graphene-based magneto-optical devices. Here we consider a system composed of two SiO₂ [27] substrates which are coated with monolayer graphene sheets in the *xy* plane and separated by a vacuum gap *d*, shown schematically in Fig. 1. The

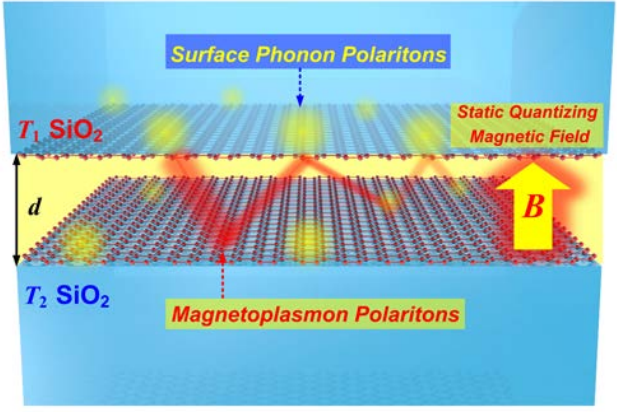

FIG. 1 Schematic of coupling between MPP and SPhPs in the NFRHT.

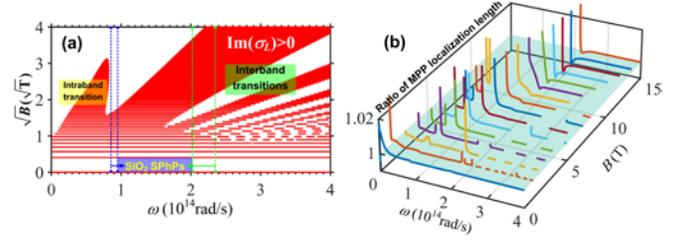

FIG. 2 (a) Imaginary part (positive region) of the longitudinal conductivities of magneto-optical graphene. (b) MPP localization length ratio of suspended graphene to graphene-coated $SiO_2$.

temperatures of the two bodies are kept at $T_1$ and $T_2$, respectively. A static quantizing magnetic field $B$ is applied perpendicularly to the system along the transverse ($z$) direction. When an external magnetic field is applied, the response of graphene electrons displays a characteristic optical quantum Hall effect [18, 28] and the conductivity becomes a tensor with nonzero elements in diagonal and off-diagonal parts

$$\begin{pmatrix} \sigma_{xx} & \sigma_{xy} \\ \sigma_{yx} & \sigma_{yy} \end{pmatrix} = \begin{pmatrix} \sigma_L & \sigma_H \\ -\sigma_H & \sigma_L \end{pmatrix} \quad (1)$$

where $\sigma_L$ and $\sigma_H$ represent the longitudinal and Hall conductivities. The magneto-optical conductivity obeys the simple form in the random phase approximation [20, 28]

$$\sigma_{L(H)}(\omega, B) = g_s g_v \times \frac{e^2}{4h} \sum_{n \neq m = -\infty}^{+\infty} \frac{\Xi_{L(H)}^{nm}}{i\Delta_{nm}} \frac{n_F(E_n) - n_F(E_m)}{\hbar\omega + \Delta_{nm} + i\Gamma_{nm}(\omega)} \quad (2)$$

where $g_{s(v)}$=2 is the spin (valley) degeneracy factor of graphene, $e$ is the charge of an electron, and $h$ is Planck's constant. $n_F(E_n) = 1/\left[1 + e^{(E_n - \mu)/k_B T}\right]$ represents the Fermi distribution function, and $k_B$ is the Boltzmann constant. $\mu$ and $T$ are the chemical potential and temperature of graphene, which are taken as $\mu$=0.06 eV and $T$=300 K throughout the letter. $\Gamma_{nm}(\omega)$ is the LL broadening, taken as 6.8 meV. $\Delta_{nm} = E_n - E_m$ is the LL energy transition ($m$, $n$= 0, ±1, ±2, … are the LL indices). The details of the magneto-optical conductivities can be found in Refs. [20, 28]. The magneto-optical conductivity of graphene is mainly determined by the longitudinal part and the MPP can be excited when Im($\sigma_L$)>0, thus the imaginary parts of $\sigma_L$ are plotted in Fig. 2(a) to reveal the characteristics of MPP. There are two types of electronic transitions: (i) intraband transitions relating the adjacent LLs ($N_F$ to $N_F$+1), where $N_F$=int[$(\mu/E_1)^2$] is the number of last occupied electron-degenerate LLs; and (ii) interband transitions. According to definition of LLs, the energy of the first LL ($E_1$) will increase with $B$ and then $N_F$ will decrease. When $N_F$ reduces to zero, the intraband transitions fade out and interband transitions dominate the magneto-optical conductivities. A series of interband transitions are observed in Fig. 2(a) and they satisfy the relation $\omega \propto \sqrt{B}$.

The decaying properties of the modes are crucial for the energy transport in the near-field regime, where the surface plasmons are always confined to the interfaces. In Fig. 2(b), the MPP localization length ratio of suspended graphene to graphene-coated $SiO_2$ are plotted to evaluate the coupled decaying properties of SPhPs to MPP. By solving the dispersion relation of modes, the wave vector satisfies [17]

$$\kappa(\omega) = \frac{X(\omega) - \sqrt{4f(\omega)g(\omega) + X(\omega)^2}}{2f(\omega)} \quad (3)$$

where $f(\omega) = \frac{i\sigma_L}{2\omega\varepsilon_0}$, $g(\omega) = i\omega\mu_0\sigma_L/2$, and $X(\omega) = -1 + f(\omega)g(\omega) - \mu_0\sigma_H^2/4\varepsilon_0$. $\varepsilon_0$ and $\mu_0$ are the permittivity and permeability of vacuum, respectively. The complex wave vector comply with the relation $q(\omega) = \sqrt{\kappa(\omega) + \varepsilon_r \omega^2/c_0^2}$. $\varepsilon_r$ denotes the relative permittivity of the medium surrounding graphene, which equals 1 and ($\varepsilon_{SiO2}$+1)/2 for suspended graphene and graphene-coated $SiO_2$, respectively. Thus, the decaying properties of the modes, i.e., the localization length in $z$ direction can be demonstrated with $\kappa'/q'$. The ratios in Fig. 2(b) imply that the coupling of SPhPs to MPP results in attenuation in localization length at some frequencies, and then the coupled modes can transport in larger length at these frequencies. At most of the frequencies, the ratios equal to unity, meaning that the coupling of SPhPs has no significant suppression on the propagation length of MPP.

To explore the coupled effect on NFRHT, in Figs. 3(a) and 3(b), the spectral radiative heat transfer coefficients in suspended graphene and graphene-coated $SiO_2$ systems are given for $d$=10 nm, and the radiative heat transfer coefficient is given by

$$h = \frac{1}{4\pi^2} \int_0^\infty \hbar\omega \frac{\partial n}{\partial T} d\omega \int_0^\infty \xi(\omega,\kappa)\kappa d\kappa \quad (4)$$

where $\hbar$ is Planck's constant divided by $2\pi$ and $n$=[exp($\hbar\omega/k_B T$)-1]$^{-1}$ denotes the mean photon occupation number. $\xi(\omega, \kappa)$ is the energy transmission coefficient [29]

$$\xi(\omega,\kappa) = \begin{cases} \mathrm{Tr}\left[\left(\mathbf{I}-\mathbf{R}_2^\dagger\mathbf{R}_2\right)\mathbf{D}\left(\mathbf{I}-\mathbf{R}_1\mathbf{R}_1^\dagger\right)\mathbf{D}^\dagger\right], \kappa<\kappa_0 \\ \mathrm{Tr}\left[\left(\mathbf{R}_2^\dagger-\mathbf{R}_2\right)\mathbf{D}\left(\mathbf{R}_1-\mathbf{R}_1^\dagger\right)\mathbf{D}^\dagger\right]e^{-2|\kappa_z|d}, \kappa>\kappa_0 \end{cases}$$

(5)

where $\kappa$, $\kappa_0=\omega/c$, $\kappa_z=\sqrt{\kappa_0^2-\kappa^2}$ are the surface-parallel wave vector, the wave vector and tangential wave vector in vacuum. $\mathbf{D}=\left(\mathbf{I}-\mathbf{R}_1\mathbf{R}_2 e^{2i\kappa_z d}\right)^{-1}$ and $\mathbf{R}$ is the reflection coefficient matrix, which can be referred in Refs. [5, 12, 21]. Above Fig. 3(b), the results for bare SiO$_2$ are demonstrated to distinguish the contribution of MPP. Comparing the results in Figs. 3(a) and 3(b), the coupling of SPhPs remoulds the energy transport of the system to a distinct regime. The interactions between graphene and SiO$_2$ are strong, especially near the frequencies where SPhPs occur, which are indicated by dashed lines. Below the low-frequency SPhPs, the consecutive branch of heat flux dominated by intraband MPP is divided into two parts near $B$=1-2 T, and the cut-off $B$ of it decays. Between the two branches of SPhPs, the flux shrinks to a narrower frequency region and higher $B$ space, which results in an intermittent region just below the high-frequency region of SPhPs. An interesting phenomenon exists in the region of the high-frequency SPhPs that the low-frequency boundary of the contour strictly coincides with the cut-off frequency of SPhPs. In addition, inside the region of high-frequency SPhPs, the branches of heat flux tend to be more compact in the space of magnetic fields.

Together with the heat transfer coefficients, the evolution of energy transmission coefficients $\zeta(\omega,\kappa)$ with magnetic fields are demonstrated in Fig. 3 to develop a deeper understanding of the coupling of SPhPs to MPP. As is known, in the absence of magnetic fields, the SPhPs of SiO$_2$ can interact strongly with graphene SPPs and push the branch of SPPs to lower frequencies [23, 25], which is also observed in Fig. 3(d). When the magnetic field is applied, the continuum energy spectrum converts into separated Landau levels in graphene and then the continuum SPPs convert into a series of separated branches of MPP. As $B$ increases, the $\zeta(\omega,\kappa)$ dominated by the intraband transitions at low frequencies gradually decay and nearly fade out when $B$ is larger than 7 T. Compared with the trend in Fig. 3(c), the coupled modes of intraband transitions deteriorate faster with $B$, which is mainly attributed to the pushing effect of SPhPs. At $B$=1-3 T, the $\zeta(\omega,\kappa)$ in Fig. 3(c) indicates that the first branch of interband MPP separates with the intraband MPP. Moreover, the results in Fig. 3(d) demonstrate that the first interband transition modes coincide with the low-frequency SPhPs and the coupling occurs. Thus, the separation phenomenon of intraband MPP observed in Fig. 3(b) can be well explained by the above evolution of $\zeta(\omega,\kappa)$. Similarly, the high-frequency SPhPs can also couple with the second branch of interband MPP. As $B$ increases, the two coupled modes both become wider and spread to higher $\omega$ and $\kappa$ space, which means the modes are becoming stronger. When $B$ is higher than 4 T, the two branches of interband transition modes continue to move towards higher $\omega$ and separate from SPhPs. Compared with the uncoupled MPP modes, the coupling effect makes the MPP modes narrower, and at the same time, they are much weaker and shrink in the direction of wave vector. As discussed above, as $B$ increases, the first branch of interband MPP moves from the low-frequency to the high-frequency SPhPs, and decay obviously at the same time. The evolution of the decaying MPP results in the intermittent region of $h(B,\omega)$ just below the high-frequency SPhPs in Fig. 3(b). At $B$=10-15 T, it should be mentioned that the big branch of modes above 2×10$^{14}$ rad/s are not the MPP modes like those in Fig. 3(c), but the coupled modes by MPP and SPhPs. It implies that MPP also has pushing effect on SPhPs, but towards higher $\omega$ which is inverse to the pushing direction of SPhPs on intraband MPP. Moreover, the strong coupled modes can also vary with the magnetic fields and benefit the tunability in NFRHT.

As kinds of surface modes, MPP and SPhPs are both confined to the interfaces and the propagation length are finite [17, 25]. Therefore, in Figs. 4(a) and 4(b), at different $d$, the coupled effects of MPP on SPhPs are demonstrated. $\eta_1=h(B)/h_{SiO_2}-1$ is defined to evaluate the modulation ability of MPP in the NFRHT. A horizontal line is labeled at $\eta_1=0$, meaning that thermal stealthy of the coated graphene can be realized at these points by applying specific magnetic fields. Another interesting phenomenon is observed that the transition from SPPs to MPP suppress the enhancement dominated by the coupling between graphene and $SiO_2$. In extreme near-field regime and strong fields, the enhancement deteriorates greatly and is essentially destroyed by the coupling of MPP when $B>8$ T. The heat transfer mechanism is completely changed from enhancement to attenuation due to the strong coupling of MPP to SPhPs. In Fig. 4(b), $\eta_2=h(B)/h(0)$, the thermal magnetoresistance ratio of that in the presence of magnetic fields to that in zero field, is plotted for different $B$. A non-monotonic trend with $d$ is observed when $B$ is between 0 and 7 T, and develops to be monotonic in larger $B$. It implies that the intensity of the coupling between MPP and SPhPs has different peaks with separation distance in the NFRHT. A relatively small value of thermal magnetoresistance ratio ($\eta_2\approx0.2$) is observed at $d=10$ nm and $B=15$ T, indicating that the strong coupling has dramatic effects on the NFRHT. Moreover, the inset in Fig. 4(b) plots the $\eta_2$ as the function of $B$, and the results demonstrate that the effect of MPP gradually fade out in larger $d$ and they are more sensitive in weak fields.

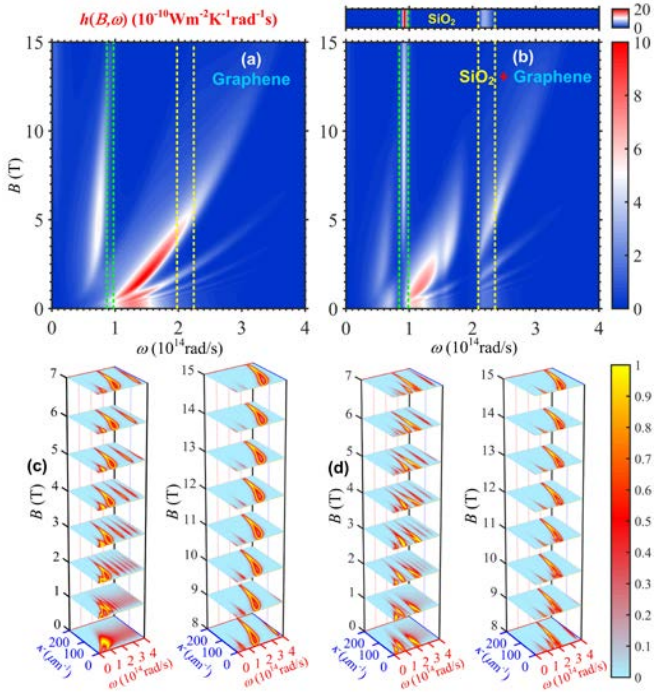

FIG. 3 For different magnetic fields, spectral radiative heat transfer coefficients in (a) suspended graphene system and (b) graphene-coated SiO$_2$ system The corresponding evolution of energy transmission coefficients are given in (c) and (d), respectively.

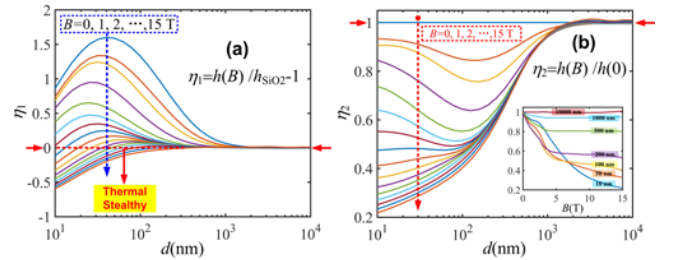

FIG. 4 At different $d$, (a) the modulation ability of MPP, and (b) the thermal magnetoresistance ratio of that in the presence of magnetic fields to that in zero field.

In conclusion, we have theoretically investigated the coupled effects of MPP and SPhPs on the NFRHT. By analyzing the decaying properties of the modes, it is found that the coupling of SPhPs to MPP may result in larger propagation length at some frequencies. With the application of magnetic field, separated branches of intraband and interband MPP can both couple with SPhPs to form strongly tunable modes. The coupling of SPhPs remoulds the energy transport of the system to a distinct regime. In extreme near-field regime and strong fields, the heat transfer mechanism is completely changed from enhancement to attenuation due to the strong coupling of MPP to SPhPs, which paves a way to realize thermal stealthy for the coated graphene. In addition, a relatively small value of thermal magnetoresistance ratio 21% is observed at $d=10$ nm and $B=15$ T, indicating that the strong coupling has dramatic effects on the NFRHT. This letter has great significance for the laboratory physics and provides practical significance for new applications in graphene-based magneto-optical devices.

**Funding.** National Natural Science Foundation of China (NSFC) (51976044, 51806047)

**Acknowledgment.** The authors acknowledge support from Heilongjiang Touyan Innovation Team Program. M. A. acknowledges support from the Institute Universitaire de France, Paris, France (UE).

**Disclosures.** The authors declare no conflicts of interest.

## References

1. D. Polder, and M. Van Hove, Physical Review B **4**, 3303-3314 (1971).
2. R. Messina, M. Antezza, and P. Ben-Abdallah, Phys Rev Lett **109**, 244302 (2012).
3. M.-J. He, H. Qi, Y.-T. Ren, Y.-J. Zhao, and M. Antezza, Applied Physics Letters **115**, 263101 (2019).
4. S. Basu, and M. Francoeur, Opt Lett **39**, 1266-1269 (2014).
5. M. He, H. Qi, Y. Ren, Y. Zhao, and M. Antezza, Optics Letters **45**, 2914 (2020).
6. X. L. Liu, and Z. M. Zhang, Applied Physics Letters **107**, 143114 (2015).
7. M.-J. He, H. Qi, Y.-F. Wang, Y.-T. Ren, W.-H. Cai, and L.-M. Ruan, Optics Express **27**, A953 (2019).
8. P. Avouris, Nano Lett **10**, 4285-4294 (2010).
9. F. J. García de Abajo, ACS Photonics **1**, 135-152 (2014).
10. T. Low, and P. Avouris, Acs Nano **8**, 1086 (2014).
11. O. Ilic, M. Jablan, J. D. Joannopoulos, I. Celanovic, H. Buljan, and M. Soljačić, Physical Review B **85** (2012).
12. M.-J. He, H. Qi, Y.-T. Ren, Y.-J. Zhao, and M. Antezza, International Journal of Heat and Mass Transfer **150**, 119305 (2020).
13. Y. Zhang, M. Antezza, H. L. Yi, and H. P. Tan, Phys. Rev. B **100**, 085426 (2019).
14. F. V. Ramirez, S. Shen, and A. J. H. McGaughey, Physical Review B **96** (2017).
15. O. Ilic, N. H. Thomas, T. Christensen, M. C. Sherrott, M. Soljacic, A. J. Minnich, O. D. Miller, and H. A. Atwater, ACS Nano **12**, 2474-2481 (2018).
16. B. Zhao, B. Guizal, Z. M. Zhang, S. Fan, and M. Antezza, Physical Review B **95** (2017).
17. A. Ferreira, N. M. R. Peres, and A. H. Castro Neto, Physical Review B **85** (2012).
18. V. P. Gusynin, and S. G. Sharapov, Physical Review B **73** (2006).
19. R. Shimano, G. Yumoto, J. Y. Yoo, R. Matsunaga, S. Tanabe, H. Hibino, T. Morimoto, and H. Aoki, Nat Commun **4**, 1841 (2013).
20. A. Ferreira, J. Viana-Gomes, Y. V. Bludov, V. Pereira, N. M. R. Peres, and A. H. Castro Neto, Physical Review B **84** (2011).
21. H. Wu, Y. Huang, L. Cui, and K. Zhu, Physical Review Applied **11** (2019).
22. L. Ge, K. Gong, Y. Cang, Y. Luo, X. Shi, and Y. Wu, Physical Review B **100** (2019).
23. M.-J. He, H. Qi, Y. Li, Y.-T. Ren, W.-H. Cai, and L.-M. Ruan, International Journal of Heat and Mass Transfer **137**, 12-19 (2019).
24. R. Messina, P. Ben-Abdallah, B. Guizal, and M. Antezza, Physical Review B **96** (2017).
25. X. L. Liu, and Z. M. Zhang, Applied Physics Letters **104**, 251911 (2014).
26. H. Yan, T. Low, W. Zhu, Y. Wu, M. Freitag, X. Li, F. Guinea, P. Avouris, and F. Xia, Nature Photonics **7**, 394-399 (2013).
27. E. D. Palik, Boston Academic Press **1**, 77-135 (1991).
28. V. P. Gusynin, S. G. Sharapov, and J. P. Carbotte, Journal of Physics: Condensed Matter **19**, 026222 (2007).
29. S. A. Biehs, P. Ben-Abdallah, F. S. Rosa, K. Joulain, and J. J. Greffet, Optics Express **19 Suppl 5**, A1088 (2011).

# Full References

1. D. Polder and M. Van Hove, "Theory of Radiative Heat Transfer between Closely Spaced Bodies," Physical Review B 4, 3303-3314 (1971).
2. R. Messina, M. Antezza, and P. Ben-Abdallah, "Three-body amplification of photon heat tunneling," Physical review letters 109, 244302 (2012).
3. M.-J. He, H. Qi, Y.-T. Ren, Y.-J. Zhao, and M. Antezza, "Graphene-based thermal repeater," Applied Physics Letters 115, 263101 (2019).
4. S. Basu and M. Francoeur, "Near-field radiative heat transfer between metamaterial thin films," Opt Lett 39, 1266-1269 (2014).
5. M. He, H. Qi, Y. Ren, Y. Zhao, and M. Antezza, "Active control of near-field radiative heat transfer by a graphene-gratings coating-twisting method," Optics Letters 45, 2914 (2020).
6. X. L. Liu and Z. M. Zhang, "Giant enhancement of nanoscale thermal radiation based on hyperbolic graphene plasmons," Applied Physics Letters 107, 143114 (2015).
7. M.-J. He, H. Qi, Y.-F. Wang, Y.-T. Ren, W.-H. Cai, and L.-M. Ruan, "Near-field radiative heat transfer in multilayered graphene system considering equilibrium temperature distribution," Optics express 27, A953 (2019).
8. P. Avouris, "Graphene: electronic and photonic properties and devices," Nano letters 10, 4285-4294 (2010).
9. F. J. García de Abajo, "Graphene Plasmonics: Challenges and Opportunities," ACS Photonics 1, 135-152 (2014).
10. T. Low and P. Avouris, "Graphene plasmonics for terahertz to mid-infrared applications," Acs Nano 8, 1086 (2014).
11. O. Ilic, M. Jablan, J. D. Joannopoulos, I. Celanovic, H. Buljan, and M. Soljačić, "Near-field thermal radiation transfer controlled by plasmons in graphene," Physical Review B 85(2012).
12. M.-J. He, H. Qi, Y.-T. Ren, Y.-J. Zhao, and M. Antezza, "Magnetoplasmonic manipulation of nanoscale thermal radiation using twisted graphene gratings," International Journal of Heat and Mass Transfer 150, 119305 (2020).
13. Y. Zhang, M. Antezza, H. L. Yi, and H. P. Tan, "Metasurface-mediated anisotropic radiative heat transfer between nanoparticles," Phys. Rev. B 100, 085426 (2019).
14. F. V. Ramirez, S. Shen, and A. J. H. McGaughey, "Near-field radiative heat transfer in graphene plasmonic nanodisk dimers," Physical Review B 96(2017).
15. O. Ilic, N. H. Thomas, T. Christensen, M. C. Sherrott, M. Soljacic, A. J. Minnich, O. D. Miller, and H. A. Atwater, "Active Radiative Thermal Switching with Graphene Plasmon Resonators," ACS Nano 12, 2474-2481 (2018).
16. B. Zhao, B. Guizal, Z. M. Zhang, S. Fan, and M. Antezza, "Near-field heat transfer between graphene/hBN multilayers," Physical Review B 95(2017).
17. A. Ferreira, N. M. R. Peres, and A. H. Castro Neto, "Confined magneto-optical waves in graphene," Physical Review B 85(2012).
18. V. P. Gusynin and S. G. Sharapov, "Transport of Dirac quasiparticles in graphene: Hall and optical conductivities," Physical Review B 73(2006).
19. R. Shimano, G. Yumoto, J. Y. Yoo, R. Matsunaga, S. Tanabe, H. Hibino, T. Morimoto, and H. Aoki, "Quantum Faraday and Kerr rotations in graphene," Nature communications 4, 1841 (2013).
20. A. Ferreira, J. Viana-Gomes, Y. V. Bludov, V. Pereira, N. M. R. Peres, and A. H. Castro Neto, "Faraday effect in graphene enclosed in an optical cavity and the equation of motion method for the study of magneto-optical transport in solids," Physical Review B 84(2011).
21. H. Wu, Y. Huang, L. Cui, and K. Zhu, "Active Magneto-Optical Control of Near-Field Radiative Heat Transfer between Graphene Sheets," Physical Review Applied 11(2019).
22. L. Ge, K. Gong, Y. Cang, Y. Luo, X. Shi, and Y. Wu, "Magnetically tunable multiband near-field radiative heat transfer between two graphene sheets," Physical Review B 100(2019).
23. M.-J. He, H. Qi, Y. Li, Y.-T. Ren, W.-H. Cai, and L.-M. Ruan, "Graphene-mediated near field thermostat based on three-body photon tunneling," International Journal of Heat and Mass Transfer 137, 12-19 (2019).
24. R. Messina, P. Ben-Abdallah, B. Guizal, and M. Antezza, "Graphene-based amplification and tuning of near-field radiative heat transfer between dissimilar polar materials," Physical Review B 96(2017).
25. X. L. Liu and Z. M. Zhang, "Graphene-assisted near-field radiative heat transfer between corrugated polar materials," Applied Physics Letters 104, 251911 (2014).
26. H. Yan, T. Low, W. Zhu, Y. Wu, M. Freitag, X. Li, F. Guinea, P. Avouris, and F. Xia, "Damping pathways of mid-infrared plasmons in graphene nanostructures," Nature Photonics 7, 394-399 (2013).
27. E. D. Palik, "Handbook of optical constants of solids II," Boston Academic Press 1, 77-135 (1991).
28. V. P. Gusynin, S. G. Sharapov, and J. P. Carbotte, "Magneto-optical conductivity in graphene," Journal of Physics: Condensed Matter 19, 026222 (2007).
29. S. A. Biehs, P. Ben-Abdallah, F. S. Rosa, K. Joulain, and J. J. Greffet, "Nanoscale heat flux between nanoporous materials," Optics express 19 Suppl 5, A1088 (2011).